\documentclass[12pt,prd,showkeys,amsmath,amssymb,nofootinbib]{revtex4-1}
\usepackage{amsmath,amsfonts,amsthm,amscd,amssymb,latexsym}
\usepackage{bm}
\usepackage{dcolumn}
\usepackage{graphicx}
\usepackage{epstopdf}
\usepackage{color}
\usepackage{epsf}
\usepackage{epsfig}
\usepackage{graphicx, epic, eepic, color}
\usepackage[colorlinks=true,urlcolor=blue,linkcolor=blue,citecolor=blue]{hyperref}

\begin{document}

\title{Modification of Aberration due to the Helicity-Rotation Coupling}

\author{Bahram \surname{Mashhoon}$^{1,2}$}
\email{mashhoonb@missouri.edu}

\affiliation{$^1$School of Astronomy, Institute for Research in Fundamental Sciences (IPM), Tehran 19395-5531, Iran\\
$^2$Department of Physics and Astronomy, University of Missouri, Columbia, Missouri 65211, USA
}

\date{\today}

\begin{abstract}
We review the physical basis for the assumption of locality in relativistic physics and its connection with the aberration of starlight. As a consequence of the hypothesis of locality, the standard relativistic formulas for the Doppler effect and aberration are independent of the polarization of the incident electromagnetic or gravitational radiation. The modification of these formulas due to the helicity-rotation coupling are discussed. In connection with the aberration of polarized radiation, we note that 
the helicity of radiation incident on a rotating observer couples to its angular velocity of rotation resulting in a slight helicity-dependent modification of the standard aberration formula. We discuss the physical origin of this effect and estimate the magnitude of the helicity aberration. 
\end{abstract}


\maketitle

%%%%%%%%%
\section{Introduction}
%%%%%%%%%

Lorentz invariance is a fundamental symmetry of nature. The Lorentz transformations connect the measurements of ideal inertial observers. These hypothetical observers remain forever at rest in inertial frames of reference that move with constant velocities with respect to each other. Inertial motion involves a rectilinear trajectory with uniform speed, in accordance with Newton's law of inertia.   Clearly, the ideal inertial observers do not exist; nevertheless, inertial physics has played a basic role in the development of physical theories~\cite{Cohen}. 

Real observers are all more or less accelerated. An accelerated observer with velocity $\bm{v}(t)$ in Minkowski spacetime may be considered locally inertial if its velocity is in effect constant during an elementary physical measurement. In this situation, the observer may employ Galilean or Lorentz transformation to connect its observation to those of the hypothetical inertial observers and hence theoretical predictions. This circumstance is illustrated in the next section, where we briefly review the wave theory of light in connection with the phenomenon of the aberration of starlight and related issues.  

To proceed decisively with the development of relativistic theory of accelerated systems, Einstein introduced the assumption that an accelerated observer with velocity $\bm{v}(t)$ in a global inertial frame is pointwise inertial; that is, the accelerated observer is locally equivalent to an otherwise identical momentarily comoving inertial observer~\cite{Einstein1, Einstein2}. To determine what an accelerated observer measures, Lorentz transformations can be applied point by point along the world line of the observer. This \emph{hypothesis of locality} provides the physical basis for the extension of the relativity theory to accelerated systems.  

Imagine an event P on the world line of an accelerated observer in a global inertial frame with Cartesian coordinates $(c\,t, \bm{x})$. For the spacetime interval along the world line at P, we can write
\begin{equation}\label{I1}
ds^2 = -c^2 dt^2 + d\bm{x}^2 = - c^2 dt^2 \left(1-\frac{v^2}{c^2}\right)\,, \qquad \bm{v} = \frac{d\bm{x}}{dt}\,,
\end{equation}
where $v = |\bm{v}|$. According to the hypothesis of locality, there is an inertial \emph{rest frame} for the observer at P with local coordinates $(c\,t', \bm{x}')$. The spacetime interval is invariant under Lorentz transformation; therefore, we have
\begin{equation}\label{I2}
ds^2 = - c^2 dt'^2\,,
\end{equation}
where the interval of local time $dt'$ corresponds to the measurement of an ideal standard clock that is carried by the accelerated observer along its world line. The standard clock displays \emph{proper time} $\tau$, $d\tau = c\,dt'$. Hence, Eqs.~\eqref{I1} and~\eqref{I2} imply
\begin{equation}\label{I3}
\tau = c\int^t  \left(1-\frac{v^2(u)}{c^2}\right)^{\frac{1}{2}}\,du\,.
\end{equation}

Let us next consider an inertial observer that stays at rest in space at $\bm{x}$  in the global inertial frame from $t_i$ to $t_f > t_i$; therefore, its proper time during this interval is $c(t_f - t_i)$.  Any accelerated observer with  $v(t) \ne 0$ that starts from 
$(c\,t_i, \bm{x})$ and returns to $(c\,t_f, \bm{x})$ will have a shorter proper time interval, since
\begin{equation}\label{I4}
\tau_f - \tau_i = c\int_{t_i}^{t_f}  \left(1-\frac{v^2(u)}{c^2}\right)^{\frac{1}{2}}\,du < c\,(t_f - t_i)\, 
\end{equation}
via Eq.~\eqref{I3}. This result forms the basis for Einstein's resolution of the twin paradox~\cite{Einstein1}. 

Using proper time, the 4-velocity of an observer can be expressed as a unit timelike vector $u^\mu = dx^\mu/d\tau$. Here, $u^\mu u_\mu = -1$ and by differentiating this relation, we note that the translational  acceleration of the observer, $A^\mu = dx^\mu / d\tau$, is orthogonal to its 4-velocity, $A^\mu u_\mu = 0$,  and is thus a spacelike 4-vector such that
\begin{equation}\label{I5}
A^\mu\,A_\mu = g^2(\tau) \geq 0\,.
\end{equation}
If the invariant magnitude of the acceleration vanishes at some instant $\tau$, i.e. $g(\tau) = 0$, the observer is locally unaccelerated; otherwise, the observer is accelerated even in its rest frame. The acceleration of the observer is in this sense absolute in the theory of relativity. The hypothesis of locality in effect replaces the accelerated observer at each instant of proper time $\tau$ by an otherwise identical momentarily comoving inertial observer. In this approximation,  the acceleration of the observer is pointwise ignored; however, it is indirectly taken into account as the pointwise application of Lorentz transformation involves the observer's velocity that keeps changing along the world line.  A critical analysis of the validity of the hypothesis of locality in relativistic physics is contained in~\cite{BMa, BMb, BMc}. The occurrence  of spin-rotation coupling reveals that the assumption of locality is valid for wave phenomena in the ray or WKB limit; moreover, it corresponds to the impulse approximation within the context of the quantum theory~\cite{BMd}. 

The pointwise application of the Lorentz transformation in the postulate of locality is based on the velocity vector $\bm{v}(t)$ that is tangent to the observer's path. At any point on a space curve, the tangent line at that point is the first approximation to the curve, while the second approximation is the osculating circle of radius $\kappa^{-1}$, where $\kappa$ is the curvature of the curve at that point. Next, the pointwise torsion of the curve is a measure of the twist of the curve out of the osculating plane containing the straight line and circle. This moving frame approach to a space curve was first developed by the French mathematicians Jean Fr\'ed\'eric Frenet in 1847 and independently Joseph Alfred Serret in 1851. It can be simply extended to the world line of an accelerated observer~\cite{BMb}. 

The locality hypothesis is an approximation; then,  what is the exact result? In this connection, it is interesting to mention here Hermann Weyl's assessment of the hypothesis of locality in the theory of relativity~\cite{Weyl}. On pages 176 and 177 of his book~\cite{Weyl}, Weyl considers  the locality assumption and concludes that this 
 approximation involves slow steady (``adiabatic") change in velocity; Weyl ends his discussion with the following sentences: 

\emph{``... The limits of acceleration within which this assumption may be made without appreciable errors arising are certainly very wide. Definite and exact statements about this point can be made only when we have built up a dynamics based on physical and mechanical laws."}

We return to the issues raised here in Section III after an exploration of the genesis of the locality hypothesis in Section II. The implications of the helicity-rotation coupling for the aberration phenomenon is discussed in Section IV.  We conclude with a brief discussion in Section V. 

%%%%%%%%%%
\section{Aberration of Starlight}
%%%%%%%%%%

The speed of light was first determined in 1676 by the Danish astronomer Ole R{\o}mer (1644-1710), who believed that Kepler's laws of planetary motion should also apply to the motion of Io around Jupiter. However, observations of the motion of Io, the closest of the Galilean moons to Jupiter, exhibited discrepancies. R{\o}mer could resolve the problems by assuming that the speed of light was finite and had a magnitude that is about $0.73$ of the accepted speed of light today. It follows from R{\o}mer's discovery that if a light signal is emitted at time $t$ and an observer that receives the signal is a distance $r$ away, the signal is received by the observer at time $t + r/c$, when the source of the signal is in general at a different position along its trajectory. R{\o}mer's discovery occurred during Isaac Newton's lifetime. Isaac Newton (1643-1727) in his optical researches formulated the corpuscular theory of light, which involved discrete light particles moving at finite speed on straight lines in accordance with Newtonian mechanics.  
R{\o}mer's idea regarding the finiteness of the speed of light was fully accepted only after the discovery of the Aberration of Starlight by James Bradley in 1728~\cite{Bradley}. 

In the process of precise determination of the celestial positions of stars, the English astronomer James Bradley (1692-1762) discovered the phenomenon of the aberration of starlight.  Bradley noticed that the positions of stars he had determined some months earlier had shifted in a systematic way. He explained this occurrence using Newton's corpuscular theory of light~\cite{Bradley}. 

Consider the global inertial frame of reference in which the Sun remains approximately at rest and the Earth revolves around the Sun with velocity $\bm{v}$. This motion occurs with an average speed of about 30 km/s. Let us assume that at some time $t$, when the light particles from a star are received by an observer on the Earth, the trajectories of the light particles and the path of the Earth on its orbit form the instantaneous $(X, Y)$ plane of a Cartesian coordinate system $(X, Y, Z)$ in the global inertial frame such that the Earth moves in the $X$ direction, see Figure 1.  Let $\bm{u}$ be the velocity of the starlight in the background global inertial frame and $\bm{u}' = \bm{u} - \bm{v}$ be its velocity as perceived by the Earth-bound observer. Alternatively, we can introduce the momentary inertial frame that moves with velocity $\bm{v}$ relative to the global inertial frame and in which the Earth is instantaneously at rest; then,  it follows from the Galilean transformation between the two inertial frames that
\begin{equation}\label{S1} 
\bm{u} = \bm{u}' + \bm{v}\,.
\end{equation}
Writing the components of these velocities with respect to the $(X, Y)$ coordinate system,
\begin{equation}\label{S2} 
u_X = \bm{u} \cdot \hat{\bm{X}} = u \cos \alpha\,, \qquad u_Y =  \bm{u} \cdot \hat{\bm{Y}} = u \sin \alpha\,
\end{equation}
and
\begin{equation}\label{S3} 
u'_X =  \bm{u}' \cdot \hat{\bm{X}} = u' \cos \alpha'\,, \qquad u'_Y =  \bm{u}' \cdot \hat{\bm{Y}} = u' \sin \alpha'\,,
\end{equation}
we find
\begin{equation}\label{S4} 
u' \cos \alpha' = u \cos \alpha - v\,, \qquad u' \sin \alpha' = u \sin \alpha\,.
\end{equation} 
Here, $\hat{\bm{X}}$ is the unit vector in the $X$ direction, etc. Moreover,   $u$, $u'$ and $v$ are the magnitudes of the velocities under consideration. Therefore, 
\begin{equation}\label{S5} 
\tan \alpha' = \frac{u \sin \alpha}{u \cos \alpha - v}\,,
\end{equation}
where $\alpha$ and $\alpha'$ characterize the trajectories of the light particles in the two inertial frames as in Figure 1, and $u$ is the speed of light $c$ in the global inertial frame in which the Sun is approximately at rest. Hence,
\begin{equation}\label{S6} 
\tan \alpha' = \frac{\sin \alpha}{\cos \alpha - \beta}\,, \qquad \beta := \frac{v}{c}\,.
\end{equation}

We define the aberration angle $\mathcal{A}$ via
\begin{equation}\label{S7} 
\mathcal{A} =  \alpha' - \alpha\,. 
\end{equation}
Then,  to first order in $\beta = v/c$,
\begin{equation}\label{S8} 
\mathcal{A} \approx \beta\, \sin \alpha\,.
\end{equation}
In the case of the motion of the Earth around the Sun, $\beta \approx 10^{-4}$. If the starlight falls perpendicularly on the Earth's velocity, $\alpha = \pi/2$ and the aberration angle is approximately equal to $\beta$; therefore, the telescope must be slightly tilted toward the motion of the Earth in order to capture the light particles from the star. More generally, as the Earth revolves around the Sun, the direction of the telescope must be continuously tilted toward the Earth's direction of motion by the relevant aberration angle in order to capture properly the starlight falling on the Earth. 

There are smaller aberrations due to other motions of the Earth. Bradley's precise position measurements  made it possible for him to measure the speed of light; indeed, his result for $c$,  about 301000 km/s, is nearly $0.4 \%$ higher than the value accepted today. 

%%%%%%%%%
\begin{figure} 
\begin{center}
\includegraphics[scale = 1]{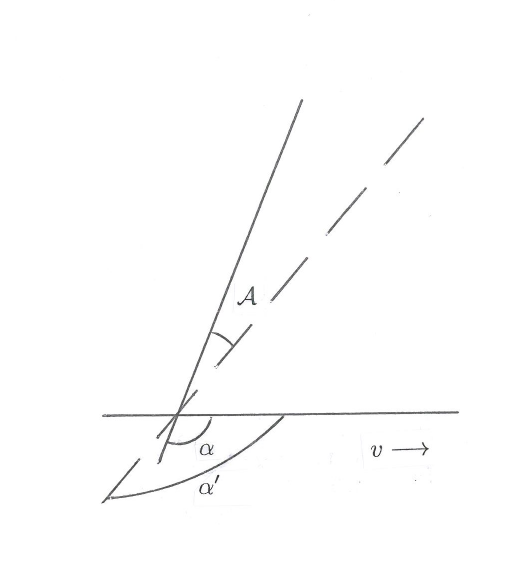}
\end{center}
\caption{Schematic illustration of the aberration angle $\mathcal{A} = \alpha' - \alpha$. The observer moves along the horizontal axis with speed $v$. The solid and dashed lines indicate the directions of incidence of light in the background and moving inertial frames, respectively. }
\end{figure}
%%%%%%%%%%%%%

Before we leave this discussion, let us note that the special relativistic formula for aberration of starlight based on Lorentz invariance is 
\begin{equation}\label{S9} 
\tan \alpha' = \frac{\sqrt{1-\beta^2}\, \sin \alpha}{ \cos \alpha - \beta}\,,
\end{equation}
which can be written as
\begin{equation}\label{S10} 
\sin \alpha' = \frac{\sqrt{1-\beta^2} \, \sin \alpha}{1-\beta \cos \alpha}\,, \quad  \cos \alpha' = \frac{\cos \alpha - \beta}{1-\beta \cos \alpha}\,, \quad \tan \frac{\alpha'}{2} = \sqrt{\frac{1+\beta}{1-\beta}}\, \tan \frac{\alpha}{2}\,.
\end{equation}
Therefore, Bradley's result is valid to first order in $\beta$. 

James Bradley succeeded in explaining the phenomenon of the aberration of starlight by recourse to Isaac Newton's corpuscular theory of light. Newton's \emph{Opticks}, published in 1704, contains the results of his optical experiments. In many experiments, Newton could explain the function of light using the particle theory. However, to explain the results of some other experiments, Newton had to assume that light exhibited wave-like properties (``theory of fits"). 

We now turn to the classical wave aspect of light. 

\subsection{Huygens-Fresnel Principle}

In 1678, the wave theory of light was proposed by Christiaan Huygens (1629-1695). The Huygens principle was further developed in 1818 by Augustin-Jean Fresnel (1788-1827). The Huygens-Fresnel principle asserts that every point on a wavefront acts as the source of spherical wavelets and there is interference between the wavelets originating from different points. The medium for the propagation of light was assumed to be the luminiferous ether. A main problem in the discussion of the results of physics experiments using the wave theory of light involved the movement of material bodies in the ether. For instance, as the Earth moved through the ether, how was the ether affected by this motion? Absence of any dragging would indicate that the ether would be completely unperturbed by the motion of the Earth. In this connection, a prominent concept was Fresnel's partial ether drag theory. Fresnel assumed that in moving through the ether, a material body carries with it the part of the ether that exists within it; otherwise, the ether is left unperturbed. 

Let $n$ be the index of refraction of a material medium such that $c/n$ would be the speed of light in this medium at rest in the ether. If the medium moves with speed $v$ in the ether and light propagates in the same direction as the motion of matter, then according to Fresnel's partial ether drag theory,  the speed of light in the moving medium is given by 
\begin{equation}\label{F1} 
V = \frac{c}{n} + v \left(1-\frac{1}{n^2}\right)\,,
\end{equation}
where $1- n^{-2}$ is the Fresnel drag coefficient and any form of dissipation is neglected.  In 1851, Armand Hippolyte Louis Fizeau (1819-1896) measured the speed of light in moving water via an interferometric arrangement. The result of Fizeau's experiment was in agreement with Eq.~\eqref{F1} and Fresnel's hypothesis.  A detailed discussion of the ether drag theories is contained in~\cite{Whittaker} and the references cited therein. 

It is evident from the Huygens-Fresnel principle that wave phenomena are in general nonlocal and this nonlocality has significant implications in connection with the theory of relativity. 

Fresnel's transverse wave theory was superseded in 1865 by Maxwell's classical wave theory of light as electromagnetic radiation. 
 
\subsection{Light is Electromagnetic Radiation}

In the course of the development of his dynamical theory of the electromagnetic field, James Clerk Maxwell (1831-1879) recognized that light is electromagnetic radiation~\cite{Maxwell}. In Maxwell's theory, a plane monochromatic linearly polarized electromagnetic wave of frequency $\omega$ and wave vector $\bm{k}$ propagating along the $z$ direction can be expressed as 
\begin{equation}\label{M1} 
\bm{E} = a \sin (\omega t - k z + \phi_0)\, \hat{\bm{x}}\,, \qquad \bm{B} = a \sin (\omega t - k z + \phi_0)\, \hat{\bm{y}}\,,
\end{equation}
where $k = \omega /c$, is the wave number, $\bm{k} = k\, \hat{\bm{z}}$,  and $\phi_0$ is a constant phase. Here, $\bm{E}$ and $\bm{B}$ are the electric and magnetic fields associated with the electromagnetic wave; moreover, each field component has an amplitude $a$ and a phase $\Phi$ defined, up to a constant, by
\begin{equation}\label{M2} 
\Phi (t, \bm{x}) = -\omega \,t + \bm{k} \cdot \bm{x}\,.
\end{equation}  

The fundamental nature of Maxwell's discovery that light is electromagnetic radiation called for the explanation of all optical phenomena known at that time in terms of electromagnetic waves. In particular, it is interesting to consider  the interpretation of the  aberration of starlight in terms of electromagnetic waves. To explain the aberration of starlight on the basis of Maxwell's theory, we consider again the global inertial frame that is approximately the rest frame of the Sun. We assume that this is also the rest frame of the luminiferous ether of Maxwell's electrodynamics such that Maxwell's equations are valid in this frame and electromagnetic waves propagate with the speed of light $c$. As before, at a given time $t$, we establish a Cartesian system of coordinates in this background inertial frame with coordinates $\bm{\mathcal{X}} = (X, Y, Z)$ such that the Earth, in its orbit around the Sun, moves along the $X$ axis with velocity $\bm{v}$ at time $t$. We are interested in the description of wave propagation in the momentary inertial frame in which the Earth is at rest. The Galilean transformation to the moving frame with Cartesian coordinates $\bm{\mathcal{X}}'$ is accomplished via 
\begin{equation}\label{M3} 
t' = t\,, \qquad \bm{\mathcal{X}}' = \bm{\mathcal{X}} - \bm{v} \,t\,.
\end{equation}  
Under this transformation, Maxwell's linear partial differential equations for $(\bm{E}, \bm{B})$ in the background inertial frame can be transformed to the moving frame. In this process,  $(\bm{E}', \bm{B}')$ will be linearly related to 
 $(\bm{E}, \bm{B})$ with coefficients that depend upon $\bm{v}$. Under such a linear transformation, the wave amplitudes will be affected, but the phase of the wave remains the same, namely,  
\begin{equation}\label{M4} 
\Phi' (t', \bm{\mathcal{X}}') = \Phi ( t, \bm{\mathcal{X}})\,,
\end{equation}  
which follows from the inspection of the wave representation such as in Eq.~\eqref{M1}. That is, under the Galilei group, 
\begin{equation}\label{M5} 
-\omega '\,t' + \bm{k}' \cdot \bm{\mathcal{X}}' =  -\omega \,t + \bm{k} \cdot \bm{\mathcal{X}}\,,
\end{equation}  
where $\omega = c |\bm{k}|$. Employing Eq.~\eqref{M3} in Eq.~\eqref{M5}, we find
\begin{equation}\label{M6} 
\omega = \omega' + \bm{v} \cdot \bm{k}'\,, \qquad \bm{k} = \bm{k}'\,.
\end{equation}   
The direction of wave propagation remains the same; therefore, no aberration occurs under a Galilean boost. However, we recover the (first-order) Doppler effect,
\begin{equation}\label{M7} 
\omega' = \omega - \bm{v} \cdot \bm{k}\,, \qquad \omega = c |\bm{k}|\,. 
\end{equation}   

Maxwell's equations of electrodynamics together with Galilean transformation fail to produce the aberration of starlight; however, we have assumed that the Sun is approximately at rest in the ether's rest frame and the Earth's motion through the ether does not perturb the ether. In this connection, Maxwell wrote to the American astronomer David P. Todd (1855-1939) regarding observational data for the eclipses of the satellites of Jupiter that could be helpful in detecting the motion of the solar system through the luminiferous ether~\cite{JCMa, JCMb, Todd}. In his letter to Todd, written in the last year of Maxwell's life, one finds, among other things, the following remark:

``..., and in the terrestrial methods of determining the velocity of light, the light comes back along the same path again, so that the velocity of the earth with respect to the ether would alter the time of the double passage by a quantity depending on the square of the ratio of the earth's velocity to that of light, and this is quite too small to be observed. ..."

Maxwell's letter was noticed by a colleague of Todd, Albert Abraham Michelson (1852-1931)~\cite{Todd}, who in 1881 designed an experiment to measure the motion of the Earth through the ether~\cite{Michelson}. An improved experiment was later performed in 1887 by Michelson and Morley~\cite{Michelson-Morley}.  We shall return to the null result of the Michelson-Morley experiment below. 

Regarding the measurement of the aberration of starlight, let us digress here and note that Newtonian point particles are intrinsically local, while waves are in general nonlocal. Consider, for instance, the motion of a point particle under the action of an external force in accordance with Newton's second law of motion. If the force vanishes at some time $t$, then the subsequent free motion of the particle takes place on a straight path in accordance with the law of inertia. Indeed, the path of the particle in the presence of the force is the envelope of the corresponding osculating straight lines. Therefore,  a classical observer passes through an infinite sequence of momentarily inertial states; however, this is only approximately valid in the presence of wave phenomena if the velocity of the observer does not change appreciably during an elementary act of measurement. This assumption underlies all of the experimental efforts  involving the wave nature of light that have been mentioned above. 

Maxwell's original theory involved linear partial differential equations for the fields $(\bm{E}, \bm{B})$ and excitations $(\bm{D}, \bm{H})$. Just as Newton's fundamental laws of motion were valid in a global inertial frame of reference, Maxwell's equations of electromagnetism were valid in a global inertial frame of reference that was at the same time the rest frame of the luminiferous ether of Maxwell's theory. In 1892, Hendrik Antoon Lorentz (1853-1928) reformulated Maxwell's theory in terms of the basic electric and magnetic fields with sources that were charged particles moving in the luminiferous ether~\cite{Whittaker}. As Whittaker remarks on page 393 of Chapter 13 of Vol. I,  ``The Classical Theories"~\cite{Whittaker}:

``The only respect in which Lorentz' medium differed from Maxwell's was in regard to the effects of the motion of bodies. Impressed by the success of Fresnel's beautiful theory of the propagation of light in moving transparent substances (footnote 1: cf. pp.108 et sqq.), Lorentz designed his equations so as to accord with that theory, and showed that this might be done by drawing a distinction between matter and aether, and assuming that a moving ponderable body cannot communicate its motion to the aether which surrounds it, or even to the aether which is entangled in its own particles; so that no part of the aether can be in motion relative to any other part. Such an aether is simply space endowed with certain dynamical properties; its introduction was the most characteristic and most valuable feature of Lorentz' theory, which differs completely from, for example, the theory of Hertz so far as concerns the electrodynamics of bodies in motion."

\subsection{Wave Explanation of the Aberration of Starlight}

In 1892, H. A. Lorentz assumed that the ether does not take part in the motion of matter~\cite{Drude}. In this way, the ether was stripped of its material nature  and was later incorporated into the spacetime structure. At this point, the failure to explain aberration of starlight only involved electrodynamics and the Galilean transformation. To account for the aberration of starlight, Lorentz suggested that the Galilean transformation needed modification. That is, Lorentz proposed a generalization of Eq.~\eqref{M3} such that 
\begin{equation}\label{L1} 
t' = t\ - \frac{\bm{v} \cdot \bm{\mathcal{X}}}{c^2}\,, \qquad \bm{\mathcal{X}}' = \bm{\mathcal{X}} - \bm{v} \,t\,.
\end{equation}  
Under this first-order Lorentz transformation, phase invariance~\eqref{M5} implies
\begin{equation}\label{L2} 
\omega = \omega' + \bm{v} \cdot \bm{k}'\, 
\end{equation}
and  
\begin{equation}\label{L3} 
\bm{k} = \bm{k}' + \frac{\bm{v}\, \omega'}{c^2}\,,
\end{equation}    
which correspond to the first-order Doppler and aberration effects in agreement with observation. To see this in detail, let us note that substituting $\bm{k}'$ from Eq.~\eqref{L3} in Eq.~\eqref{L2} and neglecting the  
$\beta^2$ term, we recover Eq.~\eqref{M7}, namely, the Doppler effect. Regarding aberration, we can write Eq.~\eqref{L3} using Eq.~\eqref{L2} as follows 
\begin{equation}\label{L4} 
\bm{k}' = \bm{k} - \frac{\bm{v}}{c^2}\,(\omega - \bm{v} \cdot \bm{k}')\,. 
\end{equation}   
Neglecting the $\beta^2$ term in this equation, we find
\begin{equation}\label{L5} 
\bm{k}' = \bm{k} - \frac{\bm{v} \,\omega}{c^2}\,. 
\end{equation}   
The direction of the starlight is determined by the direction of wave propagation, namely,   
\begin{equation}\label{L6} 
k'_X = |\bm{k}'| \cos \alpha'\,, \qquad k'_Y = |\bm{k}'| \sin \alpha'\,, 
\end{equation}   
\begin{equation}\label{L7} 
k_X = |\bm{k}| \cos \alpha\,, \qquad k_Y = |\bm{k}| \sin \alpha\,. 
\end{equation}
Writing Eq.~\eqref{L5} in terms of its components, we have
\begin{equation}\label{L8} 
k'_X = k_X - \frac{v\,\omega}{c^2} \,, \qquad  k'_Y = k_Y\,.
\end{equation}      
Therefore, 
\begin{equation}\label{L9} 
\tan \alpha' = \frac{|\bm{k}|\, \sin \alpha}{|\bm{k}|\, \cos \alpha - \frac{v\, \omega}{c^2}}\,.
\end{equation}
With $\omega = c\, |\bm{k}|$ and $\beta = v/c$, we recover the light particle formula for the aberration~\eqref{S6} using light waves, namely 
\begin{equation}\label{L10} 
\tan \alpha' = \frac{\sin \alpha}{\cos \alpha - \beta}\,.
\end{equation}
 
The wave vector, 
\begin{equation}\label{L11} 
\bm{k}'  = \left(\frac{\partial \Phi'}{\partial \bm{\mathcal{X}}'}\right)_{t'}\,, \qquad \bm{k}  = \left(\frac{\partial \Phi}{\partial \bm{\mathcal{X}}}\right)_t\,,
\end{equation}
is normal to the surface of constant phase at a given instant of time; therefore, the aberration of starlight is intimately connected with the observer dependence of the simultaneity of spatially separated events~\cite{AG}. Furthermore, let us remark that while the Galilean transformation~\eqref{M3} only involved the relative velocity $\bm{v}$, the (first-order) Lorentz transformation~\eqref{L1} contains the speed of light in the luminiferous ether, $c$,  in addition to $\bm{v}$. Indeed, we recover Galilean invariance in the limit $c \to \infty$.  Lorentz's assumption that the luminiferous ether does not partake in the motion of matter means in essence that the ether is part of the structure of space and time.  This is reflected in the appearance of the speed of light $c$ in the first-order Lorentz transformation~\eqref{L1}. 

Finally, the Lorentz-FitzGerald length contraction hypothesis could explain the null result of the Michelson-Morley experiment. The Lorentz factor $\gamma = (1-v^2/c^2)^{-\frac{1}{2}}$ was then incorporated into the first-order Lorentz transformation~\eqref{L1} by Henri Poincar\'e and Lorentz such that the resulting general Lorentz transformation connecting global frames of reference formed a group under which Maxwell's equations remained invariant~\cite{Lorentz}.  For instance, a Lorentz boost connecting two global inertial frames  can be expressed as 
\begin{equation}\label{L12} 
t' = \gamma\,\left(t\ - \frac{\bm{v} \cdot \bm{x}}{c^2}\right)\,, \qquad \bm{x}' = \bm{x} - \gamma\,\bm{v} \,t + \frac{1}{v^2}(\gamma-1)(\bm{v} \cdot \bm{x})\bm{v}\,.
\end{equation}  
Using Lorentz transformation, Eq.~\eqref{F1} can be simply obtained from the relativistic formula for the addition of velocities when terms of second order in $v$ as well as dispersion of light are neglected~\cite{Whittaker, Drude, Pauli}. The invariance of the phase under Lorentz transformation, $d\Phi = \tfrac{1}{c}\,k_\mu dx^\mu$, where $x^\mu = (ct, \bm{x})$, means that the propagation 4-vector $k^\mu = (\omega, c\,\bm{k})$ transforms under the boost~\eqref{L12} as 
\begin{equation}\label{L13} 
\omega = \gamma\,\left(\omega'\ + \bm{v} \cdot \bm{k}'\right)\,, \qquad \bm{k} = \bm{k}' + \gamma\,\frac{\bm{v}}{c^2} \,\omega' + \frac{1}{v^2}(\gamma-1)(\bm{v} \cdot \bm{k}')\bm{v}\,.
\end{equation}  
These are the standard expressions for the relativistic Doppler effect and aberration.  

\section{Beyond Locality: Helicity-Rotation Coupling}

Imagine a global inertial frame with Cartesian coordinates $(ct, x, y, z)$ in Minkowski spacetime and the class of observers that revolve uniformly in the positive sense about the $z$ axis with constant angular speed $\Omega$. Each such observer rotates in a horizontal plane with fixed $z$ on a circle of radius $r \ge 0$ with constant speed $v = r \Omega$. These observers refer their physical measurements to axes that rotate uniformly about the $z$ axis, namely, 
\begin{equation}\label{N1} 
\hat{\bm{x}}' = \hat{\bm{x}} \cos \varphi + \hat{\bm{y}} \sin \varphi \,, \qquad \hat{\bm{y}}' = -\hat{\bm{x}} \sin \varphi + \hat{\bm{y}} \cos \varphi\,, \qquad \hat{\bm{z}}' = \hat{\bm{z}}\,,
\end{equation}
where $\varphi = \Omega\,t$. 

In the background global inertial frame, the linearly polarized plane electromagnetic wave given by Eq.~\eqref{M1} is incident along the $z$ axis.  To simplify matters, we adopt the radiation gauge and consider the vector potential $A^\mu = (0, \bm{A})$ associated with the incident wave given by 
\begin{equation}\label{N2} 
\bm{A} = \frac{a}{k}\,\cos (\omega t -k z + \phi_0)\, \hat{\bm{x}}\,, 
\end{equation}
where $\nabla \cdot \bm {A} = 0$. Linearly polarized radiation represents a coherent superposition of positive and negative helicity waves with equal amplitude. 

From the viewpoint of a rotating observer, the electromagnetic field precesses in the opposite (i.e. negative) sense about the $z$ axis; that is, 
\begin{equation}\label{N3} 
\bm{A} = \frac{a}{k}\,\cos (\omega t - k z + \phi_0)\, (\hat{\bm{x}}' \cos \varphi -  \hat{\bm{y}}' \sin \varphi)\,. 
\end{equation}
From 
\begin{equation}\label{N4} 
\hat{\bm{x}}' \cos \varphi -  \hat{\bm{y}}' \sin \varphi = \frac{1}{2} (\hat{\bm{x}}' + i \hat{\bm{y}}') e^{i\,\varphi} + \frac{1}{2} (\hat{\bm{x}}' - i \hat{\bm{y}}') e^{-i\,\varphi} \,,
\end{equation}
we can write
\begin{equation}\label{N5} 
\bm{A} = \frac{a}{2\,k}\, \Re\, [(\hat{\bm{x}}' + i \hat{\bm{y}}')\, e^{-i(\omega'_{+} t - k z + \phi_0)} + (\hat{\bm{x}}' - i \hat{\bm{y}}')\,e^{-i (\omega'_{-} -k z + \phi_0)}]\,, 
\end{equation}
where $\omega'_{\pm} = (\omega \mp \Omega)$. The rotating observer perceives the incident field to be a linear superposition with equal amplitudes of a positive helicity component with frequency $\omega - \Omega$ and a negative helicity component with frequency $\omega + \Omega$. Taking time dilation into account, the measured frequencies are $\gamma (\omega \mp \Omega)$. Here, $\gamma = (1-v^2/c^2)^{- \frac{1}{2}}$ is the Lorentz factor and $\tau = t/\gamma$ is the observer's proper time such that $\varphi = \gamma \Omega \tau$.  Beyond locality, phase invariance is in general no longer valid in the transition to accelerated systems.  Indeed, the observed frequencies are different from the standard transverse Doppler effect, $\omega'_{\rm D} = \gamma \omega$, of the special theory of relativity and illustrate the spin-rotation Hamiltonian $\mathcal{H}_{SR} = -\gamma \bm{S} \cdot \bm{\Omega}$\,, where $\bm{S} = \pm \hbar \hat{\bm{k}}$ is the photon spin and $\gamma \bm{\Omega}$ is the proper angular velocity of the observer~\cite{BMe, Mashhoon:1998dm, Mashhoon:2002fq, Hauck:2003gy, AndMa}.

 The spin-rotation coupling is a general phenomenon~\cite{Mashhoon:1988zz, Mashhoon:1992zz} that has observational support~\cite{Ashby, Bah1, DSH, DDSH, DDKWLSH, Geerits:2024jdt} and is due to the inertia of intrinsic spin~\cite{Mashhoon:2005ms, BMash}.  The heuristic semi-classical  approach adopted here is in conformity with relativistic quantum theory in appropriate limits~\cite{HehlNi, Soares:1995cj, Ryder, Papini:2001un, Papini:2002cp, Lambiase:2004sm, Kiefer:2004hv, Pavlichenkov:2008tm, Obukhov:2016vvk}.  

In an incident positive (negative) helicity radiation, the electromagnetic field precesses in the positive (negative) sense with the wave frequency $\omega$ about the direction of wave propagation. If the observer rotates with angular speed $\Omega$ in the positive sense about the direction of wave propagation, the positive (negative) helicity wave appears to precess with frequency $\omega - \Omega\, ( \omega + \Omega)$ with respect to the observer. This helicity-rotation coupling phenomenon is usually called the \emph{rotational Doppler effect} in analogy with the standard linear Doppler effect~\cite{Allen, G+A, SKS, GN, Garetz, E+E, Mashhoon:2024qwj}. 

The photon helicity-rotation result $\omega'_{\pm} = \omega \mp \Omega$ has extensive observational support for $\omega \gg \Omega$. On the other hand, the result is exact and for the positive helicity incident radiation with $\omega = \Omega$ leads to the circumstance that the measured radiation field loses its temporal dependence and stands completely still with respect to the rotating observer. A similar situation occurred in the pre-relativistic Doppler formula when the observer could move with $v = c$ with the wave, a circumstance noted by Einstein in his autobiographical notes~\cite{Einstein3}. Let us note that $\omega'_{\pm} = \omega - \Omega$ has come about due to the temporally nonlocal Fourier analysis of the measured field, since the local geodesic frame of a uniformly rotating observer is stationary and invariant under translation in time. To remedy the defect associated with $\omega = \Omega$, where,  by a mere rotation, an observer can stand completely still with an electromagnetic radiation field, the past history of the accelerated observer needs to be taken into account. 

Consider an accelerated observer in Minkowski spacetime and let $\chi^{\mu}{}_{\hat \alpha}(\tau)$ be the observer's adapted tetrad frame. Here, $\tau$ is the proper time of the observer. The observer measures a background electromagnetic field, $(\bm{E}, \bm{B}) \to F_{\mu \nu}$, such that at each instant of proper time $\tau$,
\begin{equation}\label{N6}
F_{\mu \nu}\, \chi^{\mu}{}_{\hat \alpha}\, \chi^{\nu}{}_{\hat \beta} = F_{\hat \alpha \hat \beta}(\tau)\,,
\end{equation} 
is the result of the pointwise determination of the exterior electromagnetic field in accordance with the hypothesis of locality. Let $\mathcal{F}_{\hat \alpha \hat \beta}(\tau)$ be the field that is actually measured by the accelerated observer. The most general linear causal relation between $\mathcal{F}_{\hat \alpha \hat \beta}(\tau)$ and $F_{\hat \alpha \hat \beta}(\tau)$ is given by
\begin{equation}\label{N7}
\mathcal{F}_{\hat \alpha \hat \beta}(\tau) = F_{\hat \alpha \hat \beta}(\tau) + \int_{\tau_0}^{\tau} K_{\hat \alpha \hat \beta}{}^{\hat \gamma \hat \delta}(\tau, \tau')\,F_{\hat \gamma \hat \delta}(\tau') \,d\tau'\,,
\end{equation}
where $\tau_0$ is the instant at which the acceleration of the observer begins and $K(\tau, \tau')$ is the kernel of this Volterra integral relation. The unknown kernel $K$ should be determined on the basis of observational data.  
Following this approach, a nonlocal theory of accelerated systems has been developed and extended to the theory of gravitation~\cite{BMf, Mashhoon:1997qc, BMg, Mashhoon:2008vr, BMh, BMB}.    

The next section is devoted to the helicity-rotation coupling in the general case of oblique incidence of electromagnetic or gravitational radiation.  The corresponding modified aberration and Doppler effects have been considered for polarized electromagnetic radiation before~\cite{BMe, Mashhoon:2002fq}. A more comprehensive treatment of the helicity aberration is given below.

\section{Helicity Aberration}

Let us imagine that a rotating observer of the previous section measures the frequency of an incident null ray of radiation with propagation 4-vector $k^\mu = (\omega, c\,\bm{k})$. The result is the Doppler frequency $\omega'_{\rm D} = -k_\mu U^\mu$, where $U^\mu = \gamma (1, \tfrac{\bm{v}}{c})$ is the 4-velocity of the observer with $\bm{v} = \bm{\Omega} \times \bm{r}$. Hence, $\omega'_{\rm D} = \gamma (\omega -\bm{v} \cdot \bm{k})$, which can be written as $\omega'_{\rm D} = \gamma (\omega -\bm{\Omega} \cdot \bm{\ell})$, where $\bm{\ell} = \bm{r} \times \bm{k}$ such that $\hbar \bm{\ell} = \bm{L}$ is the orbital angular momentum of the null ray. 

On the other hand, for an incident plane wave of frequency $\omega$ and wave vector $\bm{k}$, we find~\cite{BMB}
\begin{equation}\label{R1}
\omega' = \gamma (\omega - M \Omega)\,,
\end{equation}
where $M$ is the total (orbital plus spin) ``magnetic" quantum number along the axis of rotation. In fact, $M = 0, \pm 1, \pm2, ...$, for scalar, vector, ... radiation, whereas $M \mp \tfrac{1}{2} = 0, \pm 1, \pm2, ...$, for a Dirac field.  In the eikonal approximation ($\omega \gg \Omega$), Eq.~\eqref{R1} reduces to
\begin{equation}\label{R2}
\omega' = \gamma ( \omega - \bm{j} \cdot \bm{\Omega})\,,
\end{equation}  
where $\hbar \bm{j} = \bm{J} = \bm{L} + \bm{S}$. In the case of a photon, for instance, $\bm{L}$ is the photon orbital angular momentum~\cite{Harwit} and $\bm{S} = \pm \hbar \hat{\bm{k}}$ is its intrinsic spin. It follows from Eq.~\eqref{R2} that intrinsic spin and orbital angular momentum couple to rotation in the same way.  

To illustrate this general result in the case of aberration, let us consider the simple case of noninertial observers $\mathcal{O}_0$ that are fixed on the $z$ axis and refer their observations to the rotating axes $(\hat{\bm{x}}', \hat{\bm{y}}', \hat{\bm{z}}')$. The tetrad frame adapted to observers $\mathcal{O}_0$ is given in $(ct, x, y, z)$ coordinates by $\lambda^{\mu}{}_{\hat \alpha}$, namely, 
\begin{equation}\label{R3}
\lambda^{\mu}{}_{\hat 0} = (1, 0, 0, 0)\,, \qquad \lambda^{\mu}{}_{\hat 1} = (0, \cos \varphi, \sin \varphi, 0)\,, 
\end{equation}
\begin{equation}\label{R4}
\lambda^{\mu}{}_{\hat 2} = (0, -\sin \varphi, \cos \varphi, 0)\,, \qquad \lambda^{\mu}{}_{\hat 3} = (0, 0, 0, 1)\,.
\end{equation}
In the case of oblique incidence of a plane wave with frequency $\omega \gg \Omega$ and wave vector $\bm{k}$, a detailed investigation reveals that observers $\mathcal{O}_0$ measure an average frequency $\omega'_0$ and wave vector $\bm{k}'_0$ given by 
\begin{equation} \label{R5}
\omega'_0 = \omega \mp s\,\hat{\bm{k}} \cdot \bm{\Omega}\,, \qquad \bm{k}'_0 = \bm{k}\,,
\end{equation}
where $s=1$ for electromagnetic radiation~\cite{BMe, Mashhoon:2002fq, Hauck:2003gy} and $s=2$ for gravitational radiation~\cite{Ramos:2006sb}.

Next, we consider observers $\mathcal{O}$ that rotate with velocity $\bm{v} = \bm{\Omega}\times\bm{r}$ about the $z$ axis and have the adapted tetrad system $e^{\mu}{}_{\hat \alpha}$ that can be determined using
  $\lambda^{\mu}{}_{\hat \alpha}$ via a Lorentz boost; that is, 
\begin{equation} \label{R6}
e^{\mu}{}_{\hat 0} = \gamma ( \lambda^{\mu}{}_{\hat 0} + \beta \lambda^{\mu}{}_{\hat 2}) = \gamma (1, -\beta \sin \varphi, \beta \cos \varphi, 0)\,,  \qquad e^{\mu}{}_{\hat 1} = \lambda^{\mu}{}_{\hat 1}\,,
\end{equation}
\begin{equation} \label{R7}
e^{\mu}{}_{\hat 2} = \gamma ( \lambda^{\mu}{}_{\hat 2} + \beta \lambda^{\mu}{}_{\hat 0}) = \gamma (\beta, -\sin \varphi,  \cos \varphi, 0)\,,  \qquad e^{\mu}{}_{\hat 3} = \lambda^{\mu}{}_{\hat 3}\,.
\end{equation}
Observers $\mathcal{O}$ and $\mathcal{O}_0$ are related by a Lorentz boost; therefore, their determinations of the frequency and wave vector of the radiation must be related via Eq.~\eqref{L13} to lowest order in our eikonal approximation scheme. Thus, 
$(\omega', c \bm{k}')$ as measured by observers $\mathcal{O}$ must be related to $(\omega'_0, c \bm{k}'_0)$ by 
\begin{equation} \label{R8}
\omega' = \gamma ( \omega'_0 - \bm{v} \cdot \bm{k}'_0)\,, \qquad \bm{k}' = \bm{k}'_0 - \gamma\,\frac{\bm{v}}{c^2} \,\omega'_0 + \frac{1}{v^2}(\gamma-1)(\bm{v} \cdot \bm{k}'_0)\bm{v}\,.
\end{equation}

Let us first note that with $\bm{v} = \bm{\Omega}\times\bm{r}$ and employing Eq.~\eqref{R5},  the modified Doppler formula in Eq.~\eqref{R8} can be written as
\begin{equation} \label{R9}
\omega' = \gamma (\omega - \bm{v} \cdot \bm{k} \mp s\,\hat{\bm{k}} \cdot \bm{\Omega}) = \gamma (\omega - \bm{j} \cdot \bm{\Omega})\,, \qquad  \bm{j} = \bm{\ell} \pm s\,\hat{\bm{k}}\,,
\end{equation}
where $\hbar \bm{j}$ is the total angular momentum in agreement with Eq.~\eqref{R2}. Here, the upper (lower) sign refers to positive (negative) helicity (electromagnetic or gravitational) radiation. Modification of the Doppler effect due to spin-rotation coupling has been worked out in detail in~\cite{Mashhoon:2002fq, AndMa}. In Doppler tracking of spacecraft, circularly polarized electromagnetic radiation is routinely employed to communicate with spacecraft and both the Earth ($\Omega_{\oplus} \approx 7 \times 10^{-5}$ rad/s) and the spacecraft rotate on their respective axes. 

Regarding the modified aberration formula, let us again consider an instantaneous $(X, Y)$ plane containing $\bm{v}$ and the incident radiation wave vector  $\bm{k}$ such that the velocity of the observer is in the direction of the horizontal axis. Then,  
\begin{equation} \label{R10}
\bm{k}'_X = \gamma \left[\bm{k}_X -\frac{v}{c^2}(\omega \mp\,s\,\hat{\bm{k}} \cdot \bm{\Omega})\right]\,,\qquad \bm{k}'_Y = \bm{k}_Y\,. 
\end{equation}
As before, 
\begin{equation} \label{R11}
\tan \alpha'_{\pm} = \frac{\gamma^{-1} \sin\alpha}{\cos \alpha -\beta\,(1\mp s\,\hat{\bm{k}} \cdot \bm{\Omega}/\omega)}\,. 
\end{equation}
To first order in $\beta$, we can write the aberration angle as
\begin{equation} \label{R12}
\mathcal{A}_{\pm}^s = \alpha'_{\pm} - \alpha \approx \mathcal{A}\, (1\mp s\,\hat{\bm{k}} \cdot \bm{\Omega}/\omega)\,, 
\end{equation}
where $\mathcal{A} \approx \beta\, \sin \alpha$ by Eq.~\eqref{S8}. Here, the upper (lower) sign refers to incident positive (negative) helicity radiation. Equation~\eqref{R12}  generalizes Eq.~\eqref{S8} by taking the helicity of the incident electromagnetic ($s=1$) or gravitational radiation ($s=2$) into account. The helicity contribution vanishes if the direction of propagation of the incident radiation is normal to the angular velocity of the observer. 

The aberration angle of polarized radiation to first order in $\beta = v/c$ consists of the standard result $\mathcal{A} \approx \beta\, \sin \alpha$ plus a helicity-dependent term that is smaller than the standard result by a factor of $\Omega/\omega \ll 1$; indeed, the observation of the helicity aberration effect appears to be beyond current measurement capabilities. For instance,  in the case of the revolution of the Earth around the Sun as well as the rotation of the Earth about its axis, and for incident electromagnetic radiation with frequency $\nu = \omega/(2\pi)$ = 1 GHz, $\beta \Omega / \omega$  is of the order of $10^{-20}$, which is far too small to be measurable at present.

\section{Discussion}

The main result of this paper is a detailed treatment of astronomical aberration and its modification as a consequence of spin-rotation coupling. The extra helicity aberration effect is too small to be significant at the present time. 

For recent applications of the spin-rotation or, equivalently, spin-vorticity coupling, see, for instance,~\cite{Mashhoon:2024wvp, Fedderke:2024ncj, Yu:2022vjn, Nozaki, YYT, AAI} and the references cited therein.

\section*{Acknowledgments}

I am grateful to Claus L\"ammerzahl for an informative discussion regarding the Fizeau experiment.

%%%%%%%%%%%%%
\appendix
%%%%%%%%%%%%%

\end{document}